\documentclass{jfm}
\def\pagelimitfooter{}
\def\absfooterflag{}

\usepackage{graphicx}
\usepackage{newtxtext}
\usepackage{newtxmath}
\usepackage{natbib}
\usepackage{hyperref}
\hypersetup{
    colorlinks = true,
    urlcolor   = blue,
    citecolor  = black,
}

\newcommand{\conv}[1]{\overline{#1}}
\newcommand{\sm}[1]{\langle #1\rangle}
\newcommand{\bb}{\bar b}

\newcommand{\zstar}{z_*}

\newcommand{\Psis}{\Psi_*}
\newcommand{\Tht}{\widetilde\Theta}
\newcommand{\Lt}{{E_A^l}}
\newcommand{\St}{{E_A^s}}

\newcommand{\PiA}{{\Pi_A}}
\newcommand{\zt}{\tilde\zeta}
\newcommand{\EA}{E_A}

\newcommand{\Nst}{N_*^2}

\newcommand{\bx}{\mathbf x}

\title{Coarse-grained local available potential energy}

\author{Jacob O. Wenegrat\aff{1}
  \corresp{\email{wenegrat@umd.edu}},
  Tom\'{a}s Chor\aff{1,2},
 \and Roy Barkan\aff{3,4}}

\affiliation{\aff{1}Department of Atmospheric and Oceanic Science, University of Maryland, College Park, MD, USA
\aff{2}atdepth MRV Inc., Cambridge, MA, USA
\aff{3}Department of Geophysics, Tel Aviv University, IL
\aff{4}Department of Atmospheric and Oceanic Sciences, University of California Los Angeles, CA, USA}
\begin{document}
\maketitle

\begin{abstract}
The available potential energy (APE) of a fluid can be defined locally in space, providing useful insights into both the energetics and dynamics of stratified flows ranging from three-dimensional turbulence to planetary scale circulations. Here we develop a framework for considering the multi-scale evolution of the local APE using a spatial filtering, or coarse-graining, approach. A scale decomposition valid generally for filters with both horizontal and vertical support is proposed, and shown to admit a construction in terms of a series of band-passed energies. The physical interpretation of the decomposition is discussed, and evolution equations for the APE at scales larger, and smaller, than the filtering scale are derived --- including the cross-scale APE flux term. These results can be paired with existing frameworks for coarse-grained kinetic energy, offering the potential for examining a complete energy cycle that accounts for conversions between both spatial scales and energy reservoirs. An illustrative example of the application of this approach to a simulation of three-dimensional Kelvin-Helmholtz instability is provided, including an exact formulation of the scale-dependent instability energy budgets, and the distribution of terms such as cross-scale transfers and APE dissipation in physical space.
\end{abstract}

\section{Introduction}
Available potential energy (APE) is defined as the difference in potential energy of a fluid's state and the minimum potential energy state the fluid could reach under a volume-preserving sorting \citep{winters_available_1995}. In this manner APE represents a fundamental quantity in understanding the energetics of stratified fluids, as it quantifies the portion of the total potential energy available to be adiabatically converted to kinetic energy \citep[KE,][]{margules_uber_1903, lorenz_available_1955}. Increases in the background potential energy (BPE, the minimum potential energy state) are also tied to the dissipative term in the APE evolution equation, providing a measure of the rate of irreversible diapycnal mixing that is widely used as a diagnostic in studies of turbulent mixing \citep{winters_available_1995, winters_diascalar_1996}. 

While the APE is often considered as a globally-integrated metric, a spatially-local definition of the APE density has also been established \citep{holliday_potential_1981, scotti_interpretation_2006, roullet_available_2009, winters_available_2013, tailleux_available_2013},
\begin{equation}\label{eq:APEdef}
    E_A(b, z, t) = \int^z_{z_*(b, t)}\left[ b_*(z', t)-b\right]\mathrm{d}z' ,
\end{equation}
where $z_*(b, t)$ is the reference height of a fluid parcel in the volume-preserving sorted minimum potential energy state \citep{winters_available_1995, tseng_mixing_2001}, and $b_*(z,t)$ is the reference buoyancy (defined implicitly such that $z_*(b_*(z,t),t) = z$). The buoyancy in (\ref{eq:APEdef}), $b(\boldsymbol{x}, t) = -g\rho_o^{-1}\rho(\boldsymbol{x},t)$, is a function of position and time, however it is held constant with respect to the integration in $z'$. The local APE density can therefore be interpreted as the path-independent work required to displace fluid parcels from their resting state to their actual position \citep[][]{tailleux_available_2013}.

Importantly, the fact that this definition enables consideration of the distribution of APE in space opens up the prospect of gaining new insights into the dynamics and mixing of stratified flows \citep{caulfield_layering_2021}. Effort in this direction includes work by \citet{molemaker_local_2010} who used spectral analysis to consider how APE is fluxed across scales through baroclinic instability and frontogenesis. \citet{scotti_diagnosing_2014}, and more recently \citet{tailleux_energetically_2025}, have also shown how the local APE density can be decomposed into mean and eddy components using Reynolds averaging. This approach was applied by \citet{zemskova_available_2015} to partition the global ocean APE into time-mean and fluctuating components.

Here we are interested in a closely related problem, that is, how to apply a spatial filtering (or `coarse-graining') approach to the APE density. This approach has a number of advantages over spectral methods or Reynolds averaging, the most significant of which is that it provides control over the {spatial} length scale where the flow is decomposed while retaining information about the location of terms in physical space that give insight into  mechanisms responsible for energy transfers. This has been shown to be particularly powerful in previous work examining cross-scale fluxes of KE, where a simple functional form for the cross-scale flux term has been established \citep{germano_turbulence_1992}. Recent applications of this include problems spanning compressible and incompressible turbulence \citep{eyink_localness_2009, aluie_compressible_2011}, magnetohydrodynamics \citep{aluie_coarse-grained_2017}, and geophysical flows such as the circulation of the ocean and atmosphere \citep{ aluie_mapping_2018, schubert_submesoscale_2020, storer_global_2022, srinivasan_forward_2023, kouhen_convective_2024}. Developing a complementary framework for understanding the evolution of the coarse-grained APE would allow  investigation of how potential energy is transferred across scales, the scale-dependent exchanges between APE and KE, and the connection of cross-scale energy transfers to diabatic processes.

Our purpose therefore is to demonstrate how coarse-graining can be applied to determine the scale-dependent evolution of the local APE density. A scale-decomposition --- valid for filters with both horizontal and vertical support --- is proposed, which separates the total APE into two energy reservoirs (above and below the filter-scale) with simple physical interpretation (\S\ref{sec:CoarseGrained}). These results are combined with the existing coarse-grained KE equation to provide a complete energy cycle accounting for transfers between both spatial scales and energy reservoirs (\S\ref{sec:EnergyCycle}), and applied to an illustrative example simulation of Kelvin-Helmholtz (KH) instability (\S\ref{sec:KH}). Section \ref{sec:discussion} briefly discusses possible alternative approaches, and conclusions are summarized in section \ref{sec:conclusions}.

\section{Coarse-grained APE}\label{sec:CoarseGrained}
\subsection{Spatial filtering}\label{sec:SpatialFiltering}
The local APE density, as defined in (\ref{eq:APEdef}), can also be written in an alternative form --- more amenable to spatial filtering --- that isolates the dependence on $b$ and $z$ in two independent terms and a cross term \citep{holliday_potential_1981}.
\begin{equation}\label{eq:APE_FY}
     E_A(b, z,t) = \Theta(b, t) +\Psi_*(z,t) - b z,
\end{equation}
where
\begin{equation}
    \Psi_*(z, t) = \int^z b_*(z', t)\,\mathrm{d}z',
\end{equation}
is, up to a constant, the hydrostatic kinematic pressure of the reference state,
and
\begin{equation}
    \Theta(b, t) = b\, z_*(b, t) - \Psi_*(z_*(b, t), t) = \int^b z_*(b', t)\,\mathrm{d}b'
\end{equation}
is its convex conjugate. 
 We note that $\Psi_*$ is convex for any stably stratified reference profile, hence $E_A\ge0$ by the Fenchel-Young inequality \citep[][]{boyd_convex_2004}, and (\ref{eq:APE_FY}) can be interpreted as a Fenchel-Young divergence. 
 
We begin by deriving a separation of scales valid for filter kernels with both horizontal and vertical extent, after which we simplify to the case of horizontal filtering alone.  
Define the filter operator
\begin{equation}\label{eq:filterdef}
    \overline{f}(\boldsymbol{x}) = \int G(\mathbf{r})f(\boldsymbol{x} + \mathbf{r})\mathrm{d}\mathbf{r}.
\end{equation}
We require that the kernel satisfy $G\ge 0$, $\int G \,\mathrm{d\mathbf{r}} = 1$, and that the filter commutes with spatial differentiation. These assumptions are
standard for the coarse-graining approach, but the last one requires care in the treatment near boundaries. Prior work has extended fields across the boundary --- e.g. zero velocity over land and zero normal-gradient for buoyancy \citep{aluie_mapping_2018, zhao_inviscid_2018, storer_global_2022} --- or through harmonic extension enforcing the boundary conditions \citep{kharghani_harmonic_2026}. Here we do not address this at length, other than to comment where relevant on aspects particular to APE.

 We also define for later notational convenience the vertical filter operator
\begin{equation}
    \langle f\rangle(z) = \int g_z(r_z)f(z + r_z)\mathrm{d}r_z,
\end{equation}
where $g_z(r_z) = \int G(\mathbf{r}_h, r_z)d\mathbf{r}_h$ with subscript $h$ denoting the horizontal vector, and $r_z$ is the parcel vertical offset relative to the filter center. For simplicity we also assume that $g_z$ is vertically symmetric, although this requirement can be relaxed if desired. If the function does not depend on horizontal position then $\overline{\,\cdot\,}\Leftrightarrow\langle\cdot\rangle$.
For both of these filter operators, fields are evaluated at the point $\boldsymbol{x} + \mathbf{r}$ unless their argument is written explicitly as a function of the filter center $\boldsymbol{x}$, in which case they are evaluated there (for example $\overline{f(a(\boldsymbol{x}), b)} = \int G(\mathbf{r})f(a(\boldsymbol{x}), b(\boldsymbol{x} + \mathbf{r}))\,\mathrm{d}\mathbf{r}$). 
For a single field this reduces to (\ref{eq:filterdef}).

We are interested in how applying a spatial filter to the total APE divides into a portion associated with the coarse, or large-scale, field, and a portion below the filter-scale (small-scale). We show below that for a horizontal-filter (denoted with $\overline{\,\cdot\,}^h$) the decomposition is the obvious one, $\overline{E_A}^h = E_A(\overline{b}^h, z, t) + [\overline{E_A}^h - E_A(\overline{b}^h, z, t)]$ such that the large-scale APE is the APE of the filtered buoyancy field and the subfilter is the remainder (\S\ref{sec:horizontal}). Vertical filtering however requires special care, unsurprising given the particular role that the direction of gravity plays in the potential energy. This can be illustrated by considering a fluid at rest (Fig. \ref{fig:fourstage}c). In this case $b=b_*(z)$, which filters to $\overline{b} = \langle b_*\rangle(z)$ such that $E_A(\overline{b}, z, t)>0$ in general when the profile is nonlinear, which would require the subfilter term (if defined as a simple remainder) to be negative. A resting fluid should carry no APE in either reservoir. 

The resolution proposed here is to measure the APE of the filtered field against the \textit{filtered} reference profile, $\langle b_*\rangle(z,t)$. This is the natural choice as it is the filtered image of the state the fluid relaxes to adiabatically (the reference profile as \textit{seen at the coarse scale}), and is therefore attainable by the filtered fluid, unlike $b_*$, which the coarse field can never reach pointwise everywhere at rest for a nonlinear reference state. Using this to define the large-scale APE reservoir, and only there, yields a two-term scale-decomposition where both terms are positive semi-definite and vanish exactly at rest. Below we first introduce the construction and its physical meaning, derive the budgets, and finally relate this to the horizontal-filtering limit.

\subsection{Definition of the scale decomposition}\label{sec:split}
In this section we define a separation of scales. It is convenient to make the reference profile an explicit argument of the APE, such that for any monotonically increasing profile $\beta_*(z,t)$ with inverse $\beta_*^{-1}(b,t)$ 
\begin{equation}
    E_A(b, z, t; \beta_*) = \int^z_{\beta_*^{-1}(b,t)} [\beta_*(z',t) - b]\mathrm{d}z' \ge 0,
\end{equation}
with positivity following as in (\ref{eq:APE_FY}). Where there is not a risk of confusion we drop the reference field argument to imply the standard definition of APE, i.e. $E_A(b,z,t) = E_A(b,z,t;b_*)$ as in (\ref{eq:APEdef}).
 
The filtered reference field, $\langle b_*\rangle(z,t)$, has an associated inverse $\tilde{z}_* = \langle b_*\rangle^{-1}$ that satisfies
\begin{equation}
    \tilde{z}_*(\langle b_*\rangle(z,t),t) = z,
\end{equation}
where the tilde notation is used here, and for several related quantities below, to indicate terms that are implicitly dependent on the filtered reference profile. We emphasize that filtering and sorting do not commute, and that $\langle b_*\rangle$ here is formed from filtering the volume-preserving sorting of the \textit{total} buoyancy field, with several alternative choices of reference state discussed in section \ref{sec:discussion}. Application of a filter with vertical extent in a vertically bounded domain requires extension of both $b$ and $b_*$ across the boundary (or a kernel that varies at the wall, which we do not consider)---a property of filtering rather than the particular scale-decomposition given below. Here we do not fix a rule to do this, but require that the extended profile is non-decreasing (so that the local APE of parcels beyond the wall is well defined and non-negative), and that the extension of a fluid at rest does not add APE. These can be accomplished using extension of the wall value (\S\ref{sec:KH}), or odd reflection about the wall value (where in both cases $b$ and $b_*$ should be treated the same way). The particular choice of extension will affect the APE of fluid parcels within approximately a filter width of the boundary (also true for coarse-grained KE), \textit{and} the scale-partition of interior fluid parcels whose $\tilde{z}_*(\overline{b},t)$ sits within approximately a filter width of the boundary (which has no KE analog). 

The filtered reference field is an average over a monotonically increasing function, and hence is itself monotonically increasing with height. 
The filtered hydrostatic pressure $\langle \Psi_* \rangle$ is therefore convex, with conjugate $\widetilde{\Theta}(b,t) = b\tilde{z}_* - \langle \Psi_*\rangle(\tilde{z}_*,t)$. 
The APE of a parcel relative to the filtered reference state is then
\begin{equation}\label{eq:tildeEa}
    E_A(b, z,t;\langle b_*\rangle) = \widetilde{\Theta}(b, t) + \langle\Psi_*\rangle(z, t) - b z \ge 0,
\end{equation}
with equality when $b = \langle b_*\rangle$, i.e. when the buoyancy is locally equal to the buoyancy of the filtered reference state. Using this, the coarse-graining separation of scales is given by
\begin{equation}
  \overline{E_A(b,z,t)}=\Lt+\St,\quad
  \Lt\equiv E_A(\overline{b},z,t;\langle b_*\rangle),\quad
  \St\equiv\overline{\EA(b,z+\widetilde{\Upsilon}(\boldsymbol{x},t),t;b_*)},
  \label{eq:split}
\end{equation}
where we have defined
\begin{equation}\label{eq:UpsilonTilde}
\widetilde{\Upsilon}(\boldsymbol{x}, t) \equiv \zt(\boldsymbol{x},t) - z,    
\end{equation}
the distance a parcel at the center of the filter kernel is displaced to reach $\zt(\bx, t) = \tilde{z}_*(\overline{b}(\bx,t), t)$, the resting level of the filtered buoyancy field in the filtered reference profile. The large scale APE, $\Lt$, is positive semi-definite by (\ref{eq:tildeEa}), and the small-scale APE is an average over $E_A\ge0$, and hence $\St\ge0$. That the sum of the scale-decomposition properly recovers the filtered total APE can be seen by writing $\St=\conv{\Theta(b,t)}+\sm{\Psis}(\zt, t)-\bb\zt-\tau(z,b)=\conv{\Theta(b, t)}-\tau(z,b)-\Tht(\bb, t)$ where we have introduced the generalized second moment \citep{germano_turbulence_1992}
\begin{equation}
    \tau(f, g) = \overline{fg} - \overline{f}\,\overline{g},
\end{equation}
such that adding $\Lt$ to $\St$ cancels
$\Tht(\bb,t)$ and recovers $\conv{\EA(b,z,t)}$.

An important aspect of the scale-decomposition given in (\ref{eq:split}) is that its terms can be built as the sum of energies in distinct bands --- desirable for a multi-scale decomposition as it allows the energies to be sensibly compared as a function of scale \citep{eyink_localness_2009}. Applying a series of filters to the APE, with scales $\ell_1<\ell_2<...$ moves energy from the large-scale APE into the small-scale APE, such that the APE at scale $\ell_n$ can be decomposed as its large scale contribution plus a sum of band-passed energies at all finer scales, each smoothed by the filter increments applied after it. The large-scale APE is thus non-increasing, and the small-scale energy non-decreasing, along any nested series of filters (appendix \ref{sec:hierarchy}).

\subsection{Physical interpretation}\label{sec:physical}
\begin{figure}
    \centering
    \includegraphics[width=1\linewidth]{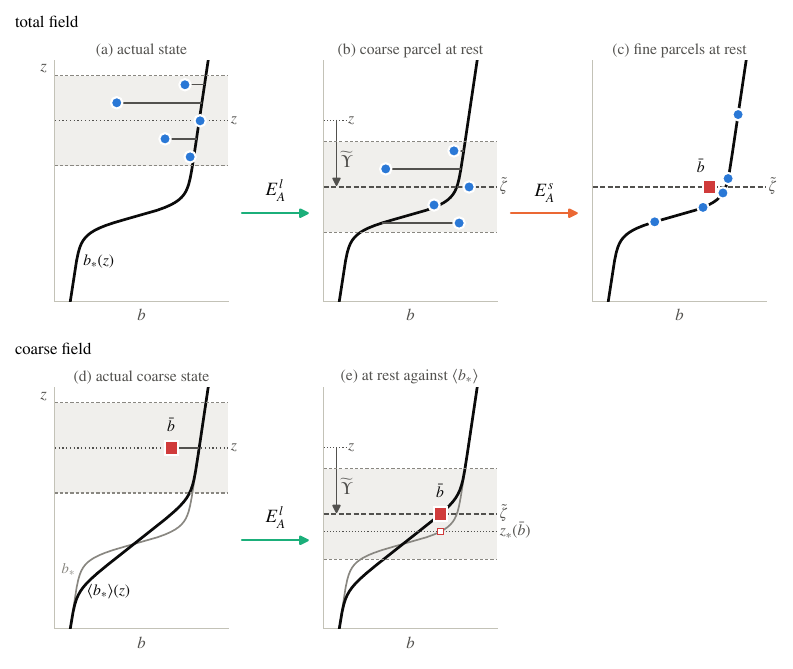}
    \caption{Schematic of the scale-decomposition for filters with vertical extent, where each row is a different view of the same process. Top row shows the total field, with fluid parcels (blue dots) that differ from the values of the sorted reference profile $b_*(z)$, such that they experience a buoyancy force $b_r$ (horizontal lines between parcels and reference profile). Bottom row shows the same adjustment process, but from the perspective of the filtered field $\overline{b}$. The shaded region indicates the filter vertical support. Release of the APE in (a) and (d) can be conceptualized as a two-stage process, beginning with release of the large-scale APE (a to b and d to e), followed by the remaining small-scale APE (b to c). Note that the coarse field at $z$ is an average over the filter (d). This field can only release APE relative to the filtered reference profile $\langle b_*\rangle(z)$ (e), and is unaffected by the release of the subfilter-scale APE (shown from b to c). Referencing the filtered field to the unfiltered profile $b_*$ would incorrectly imply the existence of large-scale APE for a profile at rest (e). In all panels the time-dependence of the reference state variables is left implicit for visual clarity.}
    \label{fig:fourstage}
\end{figure}

Physically this decomposition can be understood as dividing the work that can be released in returning the fluid to its resting state into two stages (Fig. \ref{fig:fourstage}). The first is associated with the adjustment of the coarse state to its resting position (which is determined using the filtered reference field). This adjustment releases the maximum possible APE available in the coarse field (shown below), but occurs without detailed knowledge of the subfilter arrangement of parcels. The fluid can therefore still have remaining subfilter APE, which is determined as an average over the APE of each fluid parcel under the filter kernel after they have all been displaced by the constant distance $\widetilde{\Upsilon}$ during the coarse adjustment. 

This can be seen more precisely by taking a set of fluid parcels under a filter kernel centered at height $z$ and displacing them a distance $\delta$. The work released by this shift, averaged over the filter support, is given by
\begin{equation}
    W(\delta) = \overline{\int^{z+\delta}_{z} b_r(b, z',t)\mathrm{d}z'} = \overline{b}\delta - \left[\langle\Psi_*\rangle(z+\delta, t) - \langle\Psi_*\rangle(z, t)\right],
\end{equation}
where the dependence on $b$ and $z$ is understood implicitly and
\begin{equation}\label{eq:br}
    b_r(b, z, t) \equiv b - b_*(z,t),
\end{equation} 
is the buoyancy force relative to the resting state. 
The derivative of the work with respect to $\delta$ is $\partial_\delta W = \overline{b}-\langle b_*\rangle(z+\delta)$,
the net buoyancy force on the coarse fluid, which disappears only when $\delta = \widetilde{\Upsilon}$, i.e. at the resting level of the filtered fluid in the filtered reference profile. The work released by the shift is concave with respect to the displacement, as $\partial^2_\delta W = -\langle N_*^2\rangle(z+\delta) \leqslant 0$. Thus, $\max_\delta[W(\delta)] = W(\widetilde{\Upsilon}) = E_A^l$ is the global maximum APE that can be released by relaxation of the coarse field; interpreted as adjustment to the level where the parcel has no mean buoyancy force relative to the resting state. Individual parcels under the filter kernel however can still have buoyancy anomalies relative to the reference state after the mean state adjustment, and hence APE in the form of $E_A^s$ (Fig. \ref{fig:fourstage}b).

This approach to the scale decomposition, using the filtered reference profile, can be contrasted to the potential alternative using the unfiltered reference profile, which leads to problems in decomposition. Measured against the unfiltered profile $b_*$, the natural definition of the large-scale is $E_A(\overline{b}, z,t)$ which is positive semi-definite but does not vanish at rest (Fig. \ref{fig:fourstage}c,e). As a consequence the remainder, $\overline{E_A} - E_A(\overline{b}, z,t)$, is no longer positive semi-definite (this can also be seen by noting that $E_A$ is not jointly convex in $b$ and $z$). It is possible to define a three-term split, $\overline{E_A} = \langle E_A(\overline{b}(\bx, t), z,t)\rangle + \overline{E_A(b, z_*(\overline{b}(\bx, t), t),t)} -C$, where the first two terms define positive semi-definite energies at large and small scales, respectively. This decomposition however requires the introduction of a sign-indefinite cross-term, $C = \tau(z, b)$, which is not necessarily small pointwise. Its physical origin is that vertical filtering is a mixing operator which raises the filtered fluid's center of mass (for stable stratification, Fig. \ref{fig:fourstage}e). The coarse total potential energy $ -\overline{b}z$ thus exceeds the true filtered potential energy, $-\overline{bz}$, by exactly $C$ (for a vertically centered filter kernel), such that $C=0$ only for horizontal filtering or if buoyancy and height are uncorrelated under the filter. Any decomposition that is referenced to the unfiltered background state will need to account for this lifting of the center of mass. APE is however a difference of the fluid's potential energy from the BPE, and the filtered BPE also contains a similar additional term, $-\overline{zb_*} = -z\langle b_*\rangle - \tau(z, b_*)$ which equals $C$ for a fluid at rest. Thus, defining $\Lt$ relative to the filtered reference state naturally absorbs this filter `source' of potential energy, leading to the two-term decomposition in (\ref{eq:split}). This distinction is only relevant to filters with vertical extent, as in the limit of horizontal filtering $\langle b_*\rangle = b_*$ such that the two approaches are equivalent.

The scale decomposition given by (\ref{eq:split}) thus provides a physically consistent spatial decomposition into two positive semi-definite energy reservoirs, valid for filters with both horizontal and vertical support, which we develop further below. We do however note that fluids that appear well mixed \textit{at a given scale} can still retain APE at that scale. This is somewhat contrary to the usual intuition about fluid configurations that retain APE, but is not an artifact of filtering. The resting state is set by the total fluid, such that a field which is uniform at some scale need not be in its minimum potential energy state. For example, consider two adjacent columns of fluid with buoyancies $b_1\neq b_2$. A sufficiently wide horizontal filter gives a uniform coarse field $\overline{b}$, with no large-scale variations in buoyancy. However this coarse fluid is anomalously heavy near the top and light near the bottom relative to the resting state (i.e. $\overline{b_r} \neq 0$), and therefore retains APE that can be released as the columns adjust towards rest. A similar argument can be made for a vertically well-mixed patch, which has no subfilter variance in $b$ but, once displaced to the coarse-resting level $\zt$, retains subfilter anomalies $b-b_*(z+\widetilde{\Upsilon})$ with zero mean but nonzero variance, and hence APE under the two-stage adjustment process described above (Fig. \ref{fig:fourstage}b). Thus, buoyancy variance at a given scale is neither necessary nor sufficient for there to be APE at that scale; what matters is the buoyancy anomaly relative to the rest-state appropriate to that scale.

\subsection{Evolution equations}\label{sec:evolutioneqs}
In this section we derive budget equations for each term in (\ref{eq:split}). To start, note the evolution of ${E_A}(b,z,t)$ is given by \citep[see e.g.][]{tailleux_available_2013-1}
\begin{equation}\label{eq:BasicAPEEvol}
    \frac{D E_A}{Dt} = - wb_r + \Upsilon \kappa\nabla^2b  +  R,
\end{equation}
where $\Upsilon(b, z, t) \equiv z_*(b,t) - z$ is the displacement of a parcel to its resting state in the unfiltered profile (similar to \ref{eq:UpsilonTilde}), and we assume a single-component fluid such that $Db/Dt = \kappa\nabla^2b$. The term
\begin{equation}\label{eq:RDef}
{R}(b, z,t) \equiv \dot{\Psi}_*(z,t) - \dot{\Psi}_*(z_*(b, t), t),
\end{equation}
captures the effect of time-evolution of the reference state, where the overdot notation is used to indicate the partial derivative of a reference-state function with respect to time at fixed argument (i.e. $\dot{f}(a, t) = \partial_tf|_a$). 

The budgets below follow from filtering (\ref{eq:BasicAPEEvol}), which holds inside the domain but not in general for the fields extended beyond boundaries (\S\ref{sec:split}), they are therefore exact at points whose filter kernel lies within the domain, and otherwise can carry a residual from the extension, as is also the case for coarse-grained KE.
In \S\ref{sec:KH} this residual vanishes, as the fluid at the walls is uniform and at its resting level.
Filtering (\ref{eq:BasicAPEEvol}), and writing the material derivative in terms of the filtered-flow, $D_l/Dt=\partial_t + \overline{u}_i\partial_i$ using standard Einstein summation notation with $i=(1,2,3)$ and summation over repeated indices assumed, gives
\begin{equation}\label{eq:barEa}
    \frac{D_l\overline{E_A}}{Dt} = -\overline{wb_r} + \overline{\Upsilon \kappa\nabla^2b} - \partial_i\tau(u_i, E_A) + \overline{ R}. 
\end{equation}
Forming the budget of $\Lt$ proceeds by noting that
$\partial_z\langle\Psi_*\rangle=\langle b_*\rangle$ and
$\partial_b\widetilde\Theta=\tilde z_*$. Differentiating each term in $\Lt$
(in the form given by \ref{eq:tildeEa}) following the large-scale flow gives
\begin{equation}
\frac{D_l}{Dt}\widetilde\Theta(\overline b,t)=\zt \,\frac{D_l\overline b}{Dt}
+\dot{\widetilde\Theta}(\overline b,t),
\end{equation}
\begin{equation}
\frac{D_l}{Dt}\langle\Psi_*\rangle(z,t)=\langle b_*\rangle(z,t)\,\overline w+\langle\dot\Psi_*\rangle(z,t),
\end{equation}
\begin{equation}
\frac{D_l}{Dt}\big(\overline b\,z\big)=z\,\frac{D_l\overline b}{Dt}+\overline b\,\overline w .
\end{equation}
Grouping like terms and using the filtered buoyancy equation gives
\begin{equation}
\frac{D_l\Lt}{Dt}=-\widetilde\Upsilon\,\partial_i\tau(u_i,b)
+\widetilde\Upsilon\,\kappa\nabla^2\overline b-\overline w\,\overline b_r+{ R}^l,
\end{equation}
where, from (\ref{eq:br}),
$\overline b_r = \overline b-\langle b_*\rangle(z,t)$ is the filtered buoyancy anomaly relative to the filtered reference, and
\begin{equation}
{ R}^l\equiv\langle\dot\Psi_*\rangle(z,t)-\langle\dot\Psi_*\rangle(\zt,t)
\end{equation}
captures the large-scale effect of the time-evolving reference profile. After one integration by
parts in each of the first two terms this gives
\begin{equation}\label{eq:LEvol}
\frac{D_l\Lt}{Dt}=-\overline w\,\overline b_r-\Pi_A-\partial_i F^l_i
+ { R}^l-\varepsilon^l_A.
\end{equation}
The terms on the right-hand side thus represent (from left to right) conversion between large-scale APE and KE, cross filter-scale fluxes of APE (defined below), large-scale APE transport, APE creation by reference profile evolution, and the dissipation of large-scale APE. The flux term, ${F}^l_i \equiv \widetilde{\Upsilon}({\tau}({u}_i,b) - \kappa \partial_i \overline{b})$, redistributes large-scale APE in space through small-scale advection and diffusion, but in the absence of boundary fluxes does not act as a net source or sink. The dissipation term is given by ${\varepsilon}^l_A \equiv \kappa {\partial_i}\overline{b} \partial_i \widetilde{\Upsilon}$, and can be expanded using
$\partial_i\widetilde\Upsilon=\partial_i\overline b/\widetilde{ N}_*^2( \zt,t)-\delta_{i3}$
with $\widetilde{N}_*^2=\partial_z\langle b_*\rangle$ as
\begin{equation}\label{eq:epsilonAl}
\varepsilon^l_A=\frac{\kappa|\nabla\overline b|^2}{ \widetilde{N}_*^2(\zt,t)}
-\kappa\,\partial_z\overline b ,
\end{equation}
where the first term is the sign-definite rate of increase of background potential energy due to diabatic mixing of the large-scale buoyancy field \citep[the large-scale contribution to the total $\phi_d$ in][under this decomposition]{winters_available_1995}. The second term represents conversions between internal and potential energies in the volume integral, which makes  (\ref{eq:epsilonAl}) sign indefinite \citep{tailleux_negative_2024}. This term is however a divergence, and hence can alternatively be absorbed into the transport vector if desired \citep{scotti_diagnosing_2014}, although its contribution is nonzero even in the absence of boundary fluxes. 

The cross-scale flux of APE in the coarse-graining framework is thus given by the subfilter-scale buoyancy fluxes working against the gradient of the displacement of the coarse field from its filtered reference state
\begin{equation}\label{eq:Pi_A}
    \Pi_A = - {\tau}({u}_i, b)\partial_i \widetilde{\Upsilon},
\end{equation}
such that positive values represent a transfer of APE from large to small scales. This is a primary result of this manuscript, providing the APE counterpart to the well-studied coarse-grained cross-scale kinetic energy flux term \citep[see section \ref{sec:EnergyCycle}, and][]{eyink_locality_2005, aluie_mapping_2018}. Alternatively, this can be written as,
\begin{equation} 
    \Pi_A = -\frac{\tau(u_i,b)}{\widetilde{N}_*^2(\zt,t)}\partial_i\overline{b}_r
            \;+\;\tau(w,b)\left(1-\,\frac{ \widetilde{N}_*^2(z,t)}{ \widetilde{N}_*^2(\zt,t)}\right).
\end{equation} 
This exact form can be contrasted to that which appears in the small-displacement limit, where the reference stratification is evaluated at the parcel depth, rather than the resting level, which neglects the second term on the right-hand side. In practice this is often further approximated using a horizontal- or time-mean stratification rather than the sorted profile, although this may differ substantially from $\widetilde{N}^2_*(\zt,t)$.

The evolution of the subfilter reservoir follows from $\St = \overline{E_A} - \Lt$. Subtracting (\ref{eq:LEvol}) from (\ref{eq:barEa}) gives, 
\begin{equation}
\frac{D_l\St}{Dt}=-\tau(w,b_r)+\PiA-\partial_i F^s_i+{ R}^s-\varepsilon^s_A\;
\label{eq:Sbudget}
\end{equation}
where $\tau(w, b_r)$ is the conversion of small-scale APE to small-scale KE, ${F}^s_i = {\tau}({u}_i, E_A)-\overline{\Upsilon \kappa \partial_i b} - {F}^l_i$ contains all the flux vectors that transport small-scale APE but do not flux across scale (as well as any contributions from boundary fluxes), and ${{R}}^s = \overline{{R}} - {{R}}^l$ gives the small-scale contribution of the rate of change of the reference density profile. The dissipation of APE at small-scale is defined by $\varepsilon^s_A \equiv \overline{\kappa \partial_i b \partial_i \Upsilon} -\kappa {\partial_i}\overline{b} \partial_i \widetilde{\Upsilon}$, equivalent to 
\begin{equation}
    \varepsilon^s_A=\conv{\frac{\kappa|\nabla b|^2}{\Nst(\zstar(b,t), t)}}-\frac{\kappa|\nabla\bb|^2}{\widetilde{N}_*^2(\zt, t)}.
\end{equation} 
The small-scale dissipation term thus entirely represents exchanges with the background potential energy, unlike the large-scale dissipation term which also includes a component due to conversion between internal and potential energy. Note that ${\varepsilon}_A^s$ is also not sign-definite, and can be negative, for example, in cases where the filtered reference profile is less stratified at $\zt$ than the unfiltered reference profile at $z_*(b,t)$.

It is also notable that while the term involving the rate of change of the reference profile ${R}$, (\ref{eq:RDef}), integrates to 0 over the volume \citep[][]{winters_available_1995}, this constraint only assures that $\int {{R}}^l \;\mathrm{dV} = -\int {{R}}^s \;\mathrm{dV}$ and not that either component necessarily integrates to 0, or cancels in a pointwise fashion. In this manner changes in the background state can be conceptualized as resulting in a redistribution of APE in space, which can alter the partitioning of APE between large and small scales. This is likely to be significant only in non-stationary settings where diabatic mixing affects a large fraction of the fluid such that the rate of change of $b_*$ is large (an example of which appears in section \ref{sec:KH}). 

\begin{figure}
    \centering
    \includegraphics[width=1\linewidth]{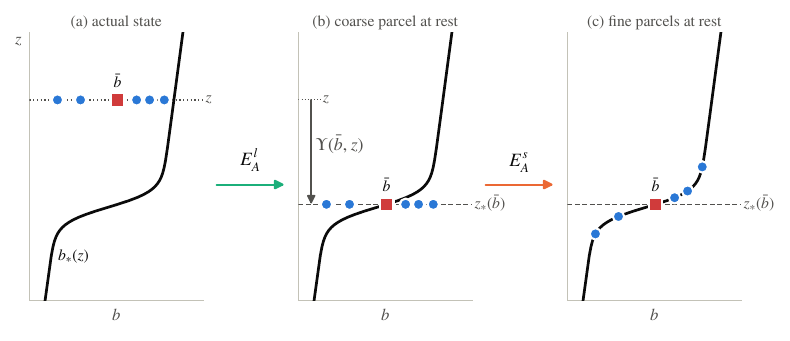}
    \caption{Schematic of the scale-decomposition for horizontal filtering. Fluid parcels within the filter region (blue dots) at height $z$ sit above their resting level (a). The large-scale APE is the portion of the APE that can be released by moving the average buoyancy field down to its resting level (b). After this adjustment, parcels under the filter have zero-mean buoyancy anomaly relative to the resting profile, but contain APE associated with the subfilter scale buoyancy anomalies (c). In all labels the superscript $h$ is dropped from the filter operator, and time-dependence of the reference state variables is left implicit, for visual clarity.}
    \label{fig:horizontal}
\end{figure}

\subsection{Horizontal filtering}\label{sec:horizontal}
In many applications, including those in geophysical flows, the primary interest is often in a horizontal scale decomposition. Here we therefore briefly consider this case, which can be understood as the limit of filter kernels with vanishing vertical extent such that $g_z = \delta(r_z)$ and $\langle f\rangle = f$. Again denoting filter kernels that operate only in the horizontal as $\overline{\,\cdot\,}^h$ the scale decomposition of (\ref{eq:split}) becomes (Fig. \ref{fig:horizontal})
\begin{equation}\label{eq:APEDecompHoriz}
    \overline{E_A(b, z, t)}^h = \underbrace{E_A(\overline{b}^h, z, t)}_{E_A^l} + \underbrace{\overline{E_A(b, z_*(\overline{b}^h(\bx), t), t)}^h}_{E_A^s}.
\end{equation}
The coarse APE is thus simply the APE of the filtered-buoyancy field, evaluated against the full reference profile, as in this case $\langle b_*\rangle(z,t) = b_*(z,t)$ (such that no extension of the reference profile is needed). The subfilter-scale APE is the average of the APE associated with each parcel over the depth range between the parcel's actual resting depth, $z_*(b, t)$, and the resting depth assigned to it by the filtered buoyancy field, $z_*(\overline{b}^h, t)$ (Fig. \ref{fig:horizontal}b). This case thus extends the Reynolds averaging approach developed in \citet{scotti_diagnosing_2014} and \citet{tailleux_energetically_2025} to operators that are not idempotent.

Budgets for each term are the same as those given for the more general case above, but with all filter operations understood to be horizontal. We note that in this case $\overline{b_r}^h = \overline{b}^h - b_*$, such that the large-scale conversion term goes as $\overline{w}^h\overline{b_r}^h = \overline{w}^h(\overline{b}^h - b_*)$, and small-scale conversions as $\tau(w, b_r) = \tau(w,b)$. Likewise the relevant displacement field of filtered parcels is $\widetilde{\Upsilon} = \Upsilon(\overline{b}^h, z, t)$, the distance between the height of the fluid and the position of the average buoyancy in the unfiltered reference profile (Fig. \ref{fig:horizontal}b).

\section{Coarse-grained energy cycle}\label{sec:EnergyCycle}
\begin{figure}
\centerline{\includegraphics[width=1\textwidth, trim=0 615 1000 25, clip=True]{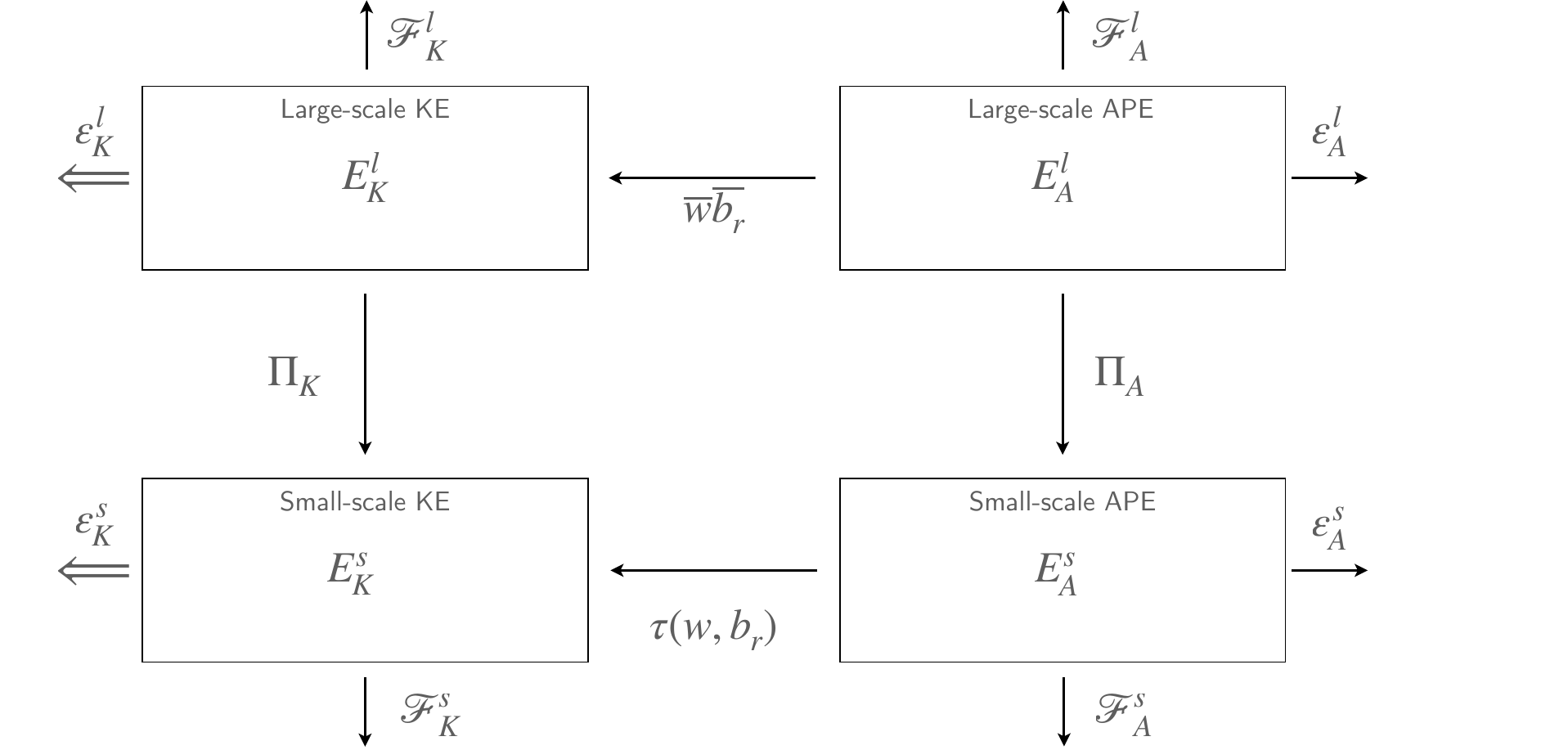}}
  \caption{Schematic energy cycle in the coarse-graining framework. Single arrows indicate the direction of sign-indefinite transfers when the term is positive, double arrows indicate sign-definite transfers.}
\label{fig:schematic}
\end{figure}
The above coarse-graining framework for APE can be combined with the existing framework for KE to form a complete energy cycle (Fig. \ref{fig:schematic}). The large- and small-scale kinetic energy are defined respectively as
\begin{equation}
    E_{K}^l = \frac{1}{2}\overline{{u}}_i\overline{{u}}_i,\quad \mathrm{and}\quad E_{K}^s = \frac{1}{2}\tau(u_i,u_i).
\end{equation}
The evolution equation for the large-scale KE is \citep{pope_turbulent_2000}
\begin{equation}\label{eq:KE_l_evol}
\frac{D_l E^l_{K}}{Dt} = \overline{w}\overline{ b}_r -\Pi_{K} -{\partial_i} {J}^l_i - \varepsilon^l_K,
\end{equation}
where $J^l_j = \overline{p}\,\overline{{u}}_j - 2\nu \overline{u}_i\overline{s}_{ij} + \overline{{u}}_i{{\tau}}({{u}_i}, {u}_j)$ gives the transport of large-scale KE due to pressure, viscous stress, and subfilter scale motions. The strain rate tensor is $s_{ij}=(\partial_ju_i + \partial_i u_j)/2$, and the dissipation of large-scale KE is $\varepsilon^l_K = 2\nu \overline{s}_{ij}\overline{s}_{ij}$. The cross-scale KE flux term is defined as \citep{aluie_mapping_2018}
\begin{equation}
    \Pi_K = -\tau(u_i,u_j)\partial_j \overline{u}_i.
\end{equation}
The small-scale KE follows a similar evolution equation \citep{eyink_localness_2009, barkan_stimulated_2017}
\begin{equation}
    \frac{D_l E_K^s}{Dt} = \tau(w, b_r) + \Pi_K - {\partial_i}{J}^s_i - \varepsilon^s_K,
\end{equation}
where $J^s_j = \tau(p,{u}_j) - 2\nu \tau(u_i,s_{ij}) + 1/2(\overline{u_ju_iu_i}-\overline{u}_j\overline{u_iu_i})-\overline{u}_i\tau(u_i, u_j)$, the terms of which have physical meaning similar to those defined following (\ref{eq:KE_l_evol}). Dissipation of small-scale KE is defined as $\varepsilon_K^s = 2\nu\,\tau( s_{ij} ,s_{ij})$. 

A distinction, relative to the standard formulation of the KE budget, is that here the pressure field should be understood to be the deviation from the hydrostatic pressure associated with the reference profile, i.e. $p = p_t - p_*$, where $p_t$ is the total pressure and $\partial p_*/\partial z = b_*$  \citep{holliday_potential_1981}. Hence the conversion terms between APE and KE (both large and small) involve $b_r$ rather than the total buoyancy $b$, although as mentioned in \S\ref{sec:horizontal} if the filter is purely horizontal then $\tau(w, b_r) = \tau(w,b)$ as the reference profile is a function only of $z$.

A schematic energy cycle is shown in Figure \ref{fig:schematic}, with subscripts indicating KE vs APE and superscripts denoting spatial scales. Contributions from boundary fluxes are represented by $\mathcal{F}$ terms, and the reader is referred to \citet{scotti_diagnosing_2014} and \citet{zemskova_available_2015} for discussion of diffusive boundary fluxes \citep[see also][]{uchoa_sink_2025}. 
 We note that this budget is open insofar as $\varepsilon_K$ and $\varepsilon_A$ are source terms for internal and background potential energy, and we have not included terms involving the rate of change of the background profile (${R}$). Alternate closed representations are possible \citep[see e.g.][]{scotti_diagnosing_2014}. 
 
\begin{figure}
    \centering
    \includegraphics[width=1\linewidth]{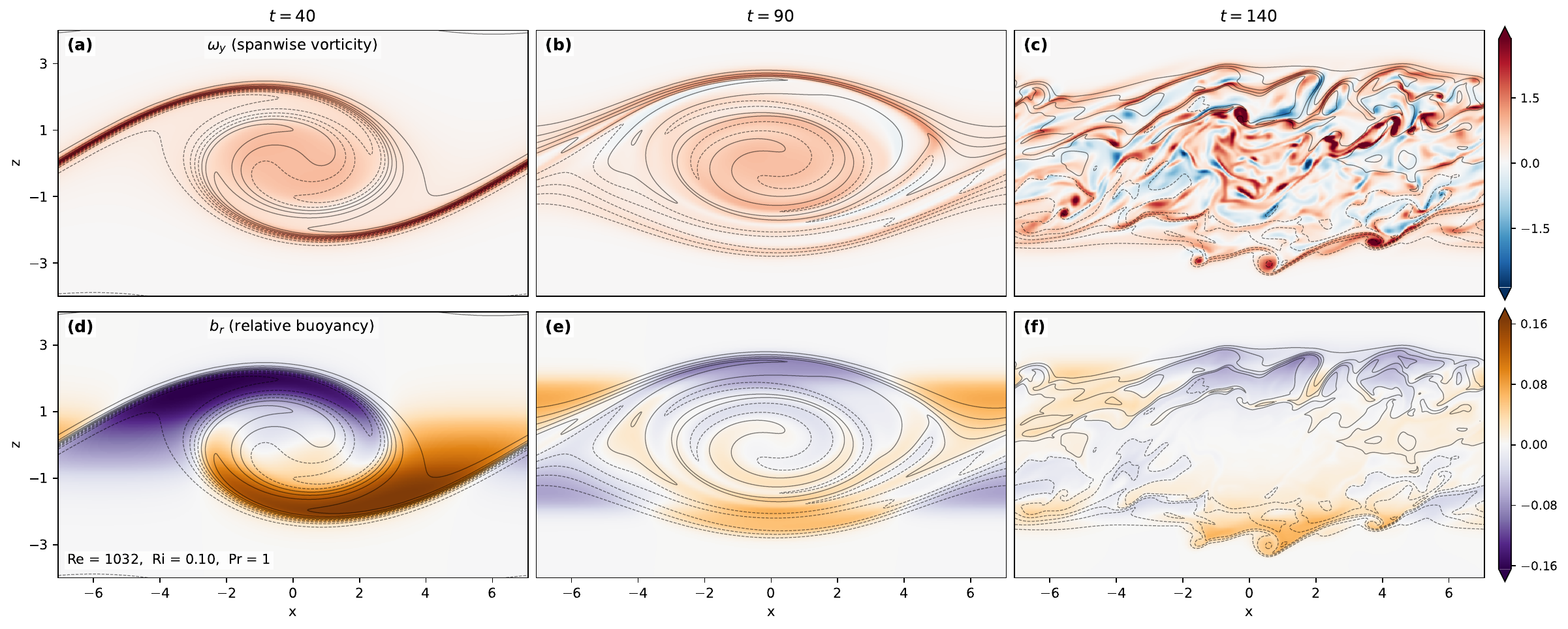}
    \caption{Snapshots of the simulated 3D Kelvin-Helmholtz instability at $y=-2.35$. Spanwise vorticity is shown in top row, and buoyancy anomalies relative to the time-evolving reference profile ($b_r = b - b_*$) are shown in the bottom row. Contours of buoyancy ($b$) are shown in all plots.}
    \label{fig:khoverview}
\end{figure}

\section{Application to Kelvin-Helmholtz instability}\label{sec:KH}
These ideas are applied here to a numerical simulation of KH instability. The simulation is performed using Oceananigans.jl \citep{wagner_high-level_2025}, a finite volume code which we use to solve the nonhydrostatic Boussinesq equations using a 4th-order centered scheme for advection and a 3rd-order Runge-Kutta scheme for time-stepping. The goal of this section is to provide a simple illustrative example, rather than an exhaustive exploration, hence we focus on a single three-dimensional simulation with $Re = Uh/\nu = 1032$, minimum initial Richardson number $Ri=0.1$, and $Pr=1$ \citep[where velocities are nondimensionalized by half the initial velocity change, $U$, and length scales by half the initial shear layer thickness, $h$, following][]{kaminski_stratified_2019}. The domain size $(L_x, L_y, L_z) = (14, 4.7,25)$ is chosen to contain 1 wavelength of the fastest growing linear mode (Fig. \ref{fig:khoverview}), which excludes the possible later development of vortex pairing \citep{winant_vortex_1974}. Discretization of ($N_x$, $N_y$, $N_z$) = (576, 192, 1024) was found to be sufficient to accurately resolve the turbulence by requiring $\Delta x$ to be smaller than 1.2 the Kolmogorov scale \citep[a stricter criterion than the factor of 2.5 used in][to account for our finite volume discretization] {kaminski_stratified_2019}. Filtering is implemented using an isotropic Gaussian kernel in $(x,y,z)$, and we identify the filter-scale $\ell$ as the full width at half maximum \citep[see][for a discussion of kernel choices]{aluie_mapping_2018}. Vertical extension of the velocity and $b$ fields,  and $b_*$ profile, is accomplished by simple padding of the wall values, equivalent for this problem to running with a vertically larger domain.

\begin{figure}
    \centering
    \includegraphics[width=1\linewidth]{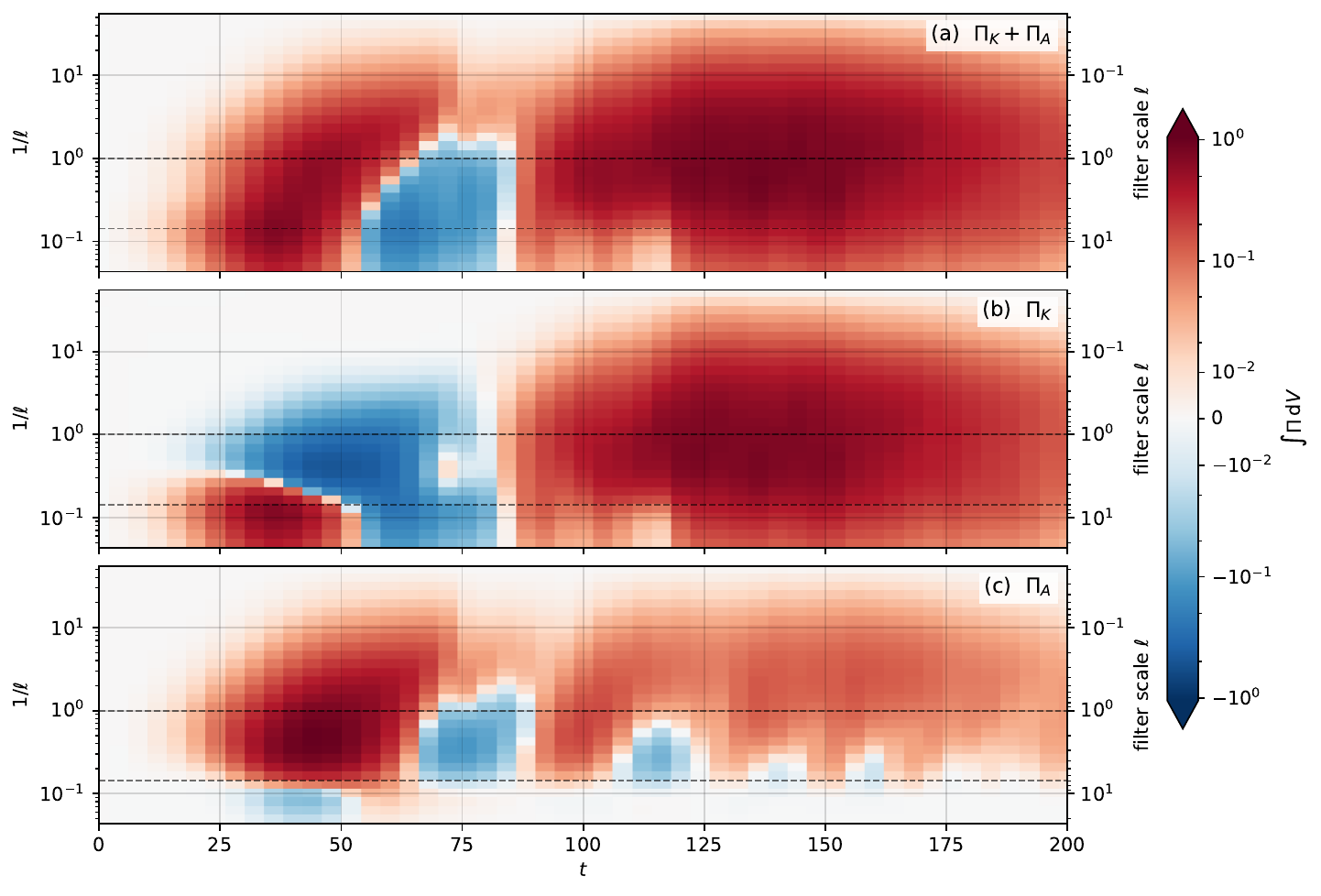}
    \caption{Cross-scale fluxes of total, kinetic, and available potential energies (subplots a-c, respectively) as a function of time and filter-scale. Dashed lines in each panel indicate scales shown in figure \ref{fig:budgets}.}
    \label{fig:sweeps}
\end{figure}

The simulated KH instability follows a standard evolution \citep[as described for instance by][]{caulfield_anatomy_2000}, beginning with a two-dimensional overturning stage ($t\approx0-90$), followed by the development of secondary spanwise overturning instabilities which then transition to turbulence (Fig. \ref{fig:khoverview}). The scale-dependent energy transfers during the instability development are shown as a function of time and filter scale in Figure \ref{fig:sweeps}. During the initial KH growth ($t\approx0-50$) there is a forward transfer of KE at large-scales ($\Pi_K>0$), associated with the growth of the instability from the horizontally uniform initial velocity field. $\Pi_K$ switches sign to negative (upscale transfer) at smaller-scales, indicating a convergence of cross-scale KE fluxes between $\ell\approx2-10$. This same range of scales has positive APE transfers, $\Pi_A>0$, largely located in the instability braids (as shown below). At about $t=50$, when buoyancy anomalies have been wound around the billow core, the total cross-scale energy transfer becomes negative for scales $\ell\gtrapprox1$, first driven by inverse KE transfers and later with an additional contribution from inverse APE transfers.

The cross-scale transfers change abruptly at the onset of secondary instabilities at $t\approx90$, when the total energy flux becomes positive at all scales. This signal is dominated by the change in sign of $\Pi_K$, which increases in magnitude through $t\approx 125$, particularly at scales $\ell\sim 1$ roughly consistent with the scale of the developing spanwise instabilities. During this period, cross-scale APE transfers are much weaker, with the most prominent signal being an oscillatory cross-scale exchange migrating towards larger scales. This begins around the time when the billow fills the numerical domain, and has an initial timescale $\approx 50$, consistent with the estimated turn-over timescale of the outer billow. Inspection of the solutions suggests this exchange is associated with nutation of the vortex \citep[see supplementary animation, and][]{klaassen_evolution_1985}. 

\begin{figure}
\centerline{\includegraphics[width=1\textwidth, trim=0 0 0 0, clip=True]{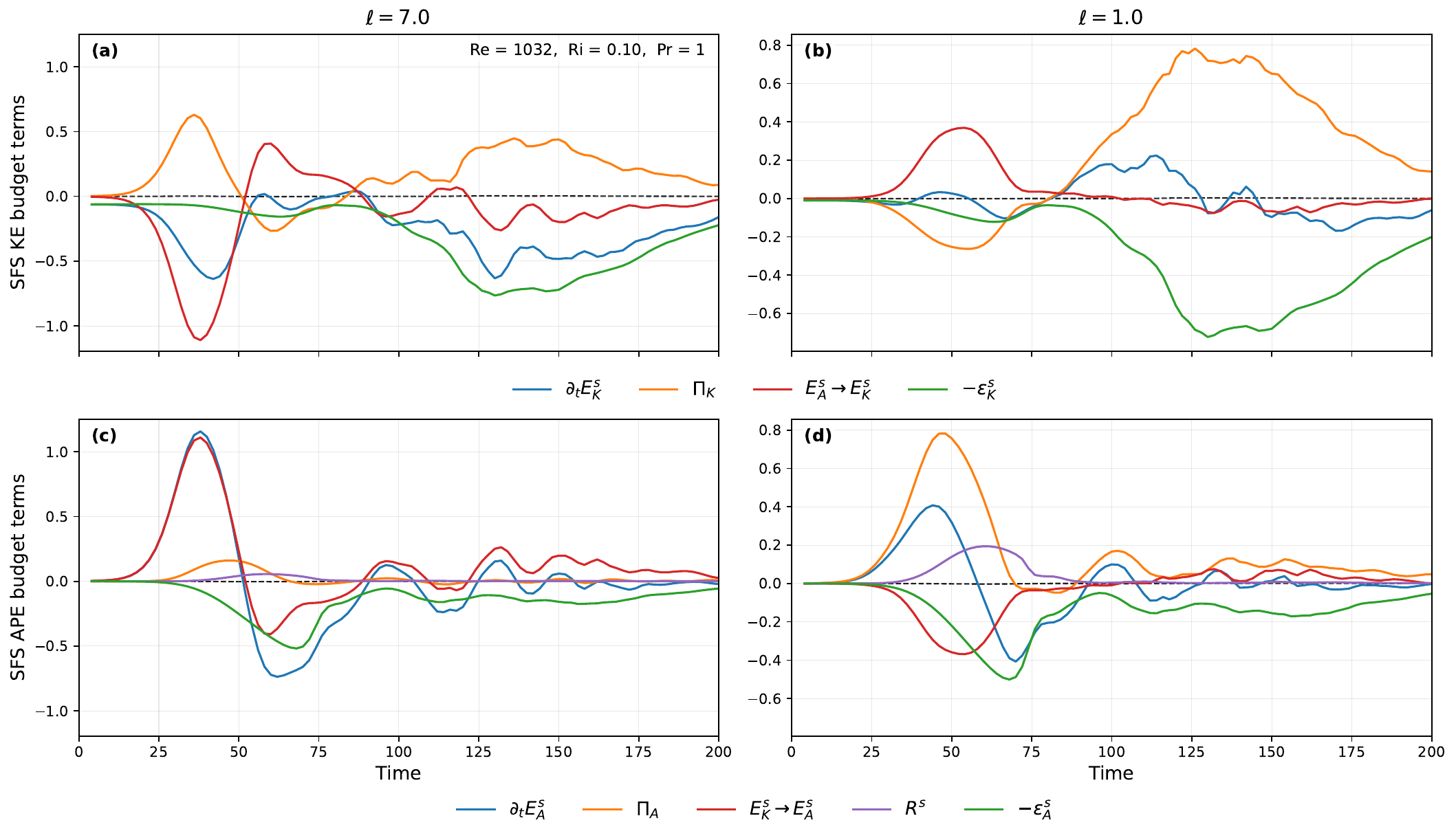}}
  \caption{Domain-integrated subfilter-scale KE (top row) and subfilter-scale APE (bottom row) energy budgets for two selected filter scales (columns, as labeled). Budget residuals are small, indicated by the black dashed lines. }
\label{fig:budgets}
\end{figure}

Time-series of the sub-filter scale KE and APE budgets are shown for two selected filter scales (Fig. \ref{fig:budgets}). At half the wavelength of the linear instability ($\ell=7$, effectively spanwise averaged for this domain) small-scale APE begins to increase around $t=20$ via conversion from KE, offset in part in the KE budget by the initially forward energy transfers. This leads in the APE budget to an increase of $E_A^s$, with a small additional contribution from forward APE transfers that peaks around $t=50$. After this time, conversion of APE-KE changes sign ($\tau(w, b_r)>0$), reducing small-scale APE, and providing a source of $E_K^s$, a portion of which is then sent back to the large scales ($\Pi_K < 0$). At the smaller scale of $\ell=1$ (half the thickness of the initial shear layer) $E_A^s$ increases until $t\approx 60$, largely due to a downscale transfer of APE ($\Pi_A>0$). This is partially offset by  conversion to KE, which is balanced in the $E_K^s$ budget primarily by an inverse transfer of KE to larger scales. This progression is consistent with the expectation that when the flow is two-dimensional there will not be an efficient direct transfer of KE to small scales, and instead small-scale KE is generated via a forward APE transfer (and then conversion) of APE to KE \citep{staquet_two-dimensional_1995}.

\begin{figure}
\centerline{\includegraphics[width=1\textwidth, trim=0 0 0 0, clip=True]{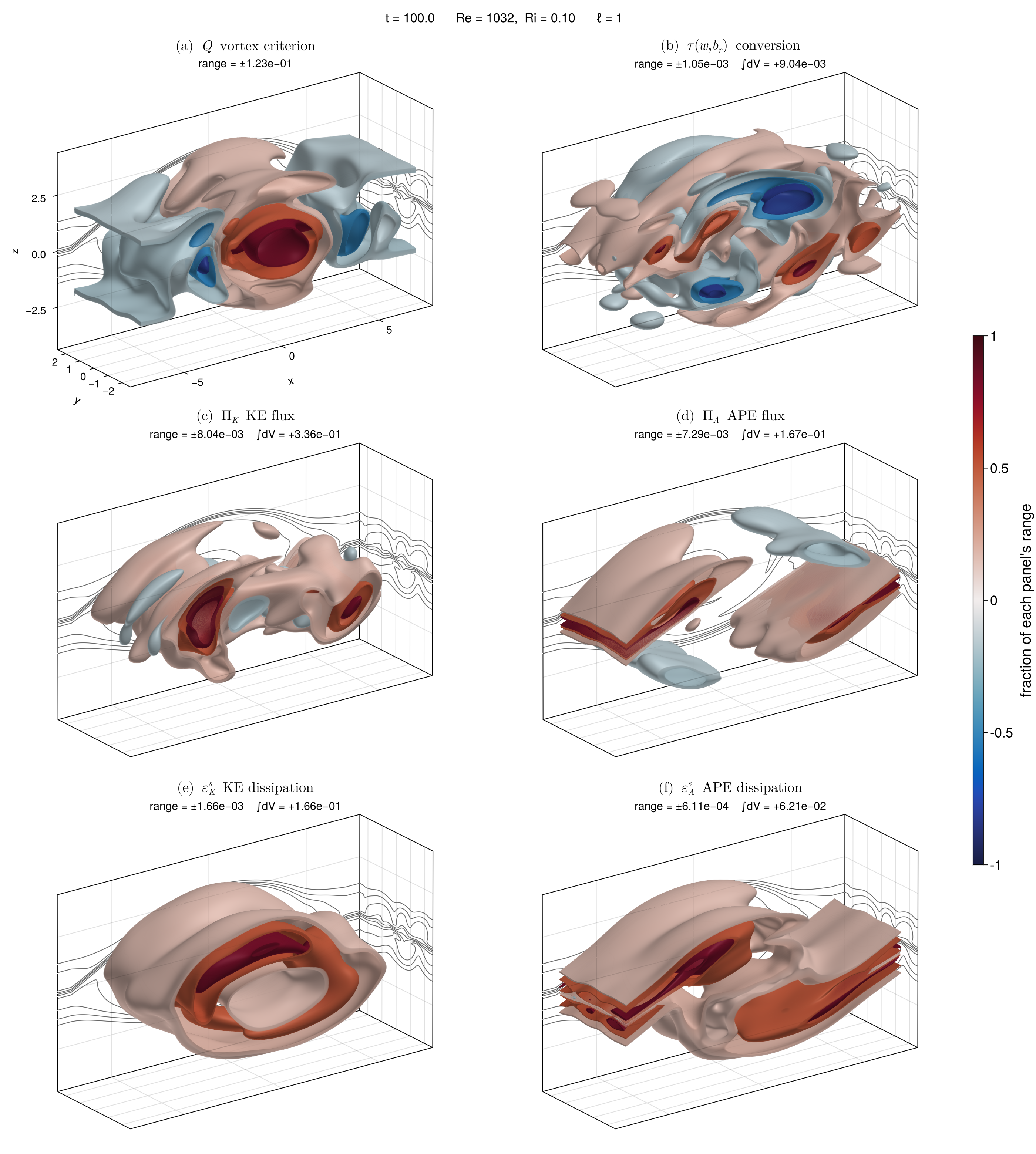}}
  \caption{Isosurface plots at $t=100$ for $\ell=1$. The vortex criterion is defined as $Q = 1/2(||\mathbf{\Omega}^l||^2 - ||\mathbf{s}^l||^2)$ where $\mathbf{\Omega}^l$ is the rotation tensor and $\mathbf{s}^l$ the strain tensor, both formed from the filtered velocity field.
 Colors are normalized in each panel by the subplot range of values (as shown in panel titles), and surfaces are plotted at $\pm0.18$, $\pm0.515$, $\pm0.85$. The domain integrated values are also given in each subplot title. Contours of buoyancy are also shown, projected against the walls, for reference. An animation of these fields is provided in the supplementary material. }
\label{fig:snapshots}
\end{figure}

At both scales the dissipation of APE peaks at approximately $t=70$ when the billow has reached the horizontal extent of the numerical domain, well after the cross-scale APE transfers have subsided. This mixing leads to a rapid change in the reference density profile, reflected in the positive contribution of ${R}^s$ to the small-scale APE budget. During this period APE-KE conversions are small in the volume-integral, although this obscures compensating dipoles of large positive and negative conversions around the core (Fig. \ref{fig:snapshots}b, discussed below). After the onset of secondary instabilities, dissipation of kinetic energy grows, balancing the increased forward KE transfers. During this late evolution at $\ell=1$ the primary source of small-scale APE is the cross-scale transfer term, with secondary contributions from positive KE-APE conversions. These sources are balanced on average by $\varepsilon_A^s$, which remains approximately steady at about 1/3 of its peak value, while the sources and tendency $\partial_tE_A^s$ vary considerably in time, suggesting the dissipation of APE is set by the accumulated subfilter anomalies rather than the instantaneous transfers.

Finally, we consider a snapshot of the spatial distribution of a subset of the energy terms for $\ell=1$ at $t=100$, shortly after the onset of secondary instabilities (Fig. \ref{fig:snapshots}). Cross-scale APE fluxes are most strongly positive along the braids, with little spanwise variability (consistent with their being driven largely by the initial two-dimensional KH instability). The pattern of APE to KE conversion shows a dipole structure around the billow core, but as noted above is small in the volume-integral after $t=75$ (Fig. \ref{fig:budgets}), suggesting little net contribution of small-scale APE conversion to the growth of the secondary instabilities \citep{caulfield_anatomy_2000}. Cross-scale KE transfers are forward, and particularly enhanced around the stagnation points of the vortex where the strain is large \citep{mashayek_zoo_2012}, although with significant spanwise structure that evolves quickly in time (see supplementary animation). Kinetic energy dissipation is enhanced along the billow eyelids \citep{caulfield_anatomy_2000}, but interestingly shows relatively little spanwise structure at this scale compared to the KE production terms ($\Pi_K$ and $\tau(w,b_r)$). APE dissipation is enhanced along the braids, particularly near the braid-billow junctions, with a small-region of transient negative small-scale APE dissipation along the centerline of the braid. This pattern is robust through the entire evolution up until the onset of fully three-dimensional turbulence (see supplementary animation), implying locally efficient mixing persists even at times where the bulk energetics favor dissipation of kinetic energy. These findings are consistent with inferences from prior observational and modeling work at higher Reynolds number using approximate forms of APE dissipation based on scalar variance dissipation \citep[][]{ geyer_mixing_2010, lewin_influence_2021, lefauve_structure_2026}, shown here using the exact scale-dependent form of $\varepsilon_A^s$. 

\section{Discussion of alternative approaches}\label{sec:discussion}
The scale-decomposition developed here is formed by coarse-graining the full local APE, (\ref{eq:APEdef}). This is advantageous in that it retains connections to the exact energetics of the fluid, but comes at the cost of additional complexity associated with filtering. This is particularly the case for filters with vertical support, which require thoughtful definition of the large-scale APE, and careful treatment of the boundaries (\S\ref{sec:split}). Here we therefore briefly review two possible alternative approaches. 

As noted above the APE can be defined relative to any reference profile that increases monotonically with height, hence one admissible option would be to define the reference state based on sorting the filtered buoyancy field, $\operatorname{sort}(\overline{b})$, rather than the total buoyancy (emphasizing that sorting the filtered buoyancy is not equivalent to filtering the sorted buoyancy). A coarse APE defined in this way, $E_A(\overline{b}, z, t; \operatorname{sort}(\overline{b}))$, may be conceptually useful as it represents the APE that can be determined without any knowledge whatsoever of the subfilter fields (the quantity for instance that is accessible in a coarsely-resolved model). However, this quantity is not always smaller pointwise than the filtered total APE, such that defining a subfilter energy becomes challenging. Instead the cross-scale flux terms can be conceptualized as an exchange directly between the large-scale APE and the sorted filtered BPE, with the reference state evolving due to adiabatic processes \citep{scotti_diagnosing_2014, venaille_statistical_2017}. Likewise, the reference sorted from $\overline{b}$ is recomputed at every scale, an operation which does not commute with filtering, hence the exact band composition (appendix \ref{sec:hierarchy}) is lost. These limitations obscure connections between cross-scale energy transfers and diabatic mixing, and thus we believe the approach used here is generally preferable.

Alternatively, a distinct class of decomposition references the small- and large-scale reservoirs to different reference states. \citet{craske_decomposition_2021} 
partitions the domain into subdomains, assigning each an `inner' APE found by sorting within the subdomain and an `outer' APE accounting for context provided by the parent domain (counterparts to our small- and large-scales). Every reference state is then an exact volume-preserving rearrangement of fluid parcels, whereas filtering is a diabatic operator (the source of much of the complexity for vertical-filtering in \S\ref{sec:CoarseGrained}). The inner-outer densities however sum to the local APE only under integration over the subdomain, so the decomposition itself, and hence its budgets, are integral rather than pointwise. Following from this, the results depend on where the partition boundary is placed, and no inner-outer decomposition of the KE budget is yet available. More significantly, inner and outer APE are exchanged through advective fluxes across subdomain boundaries, such that an adiabatic uniform velocity field generates inner-outer conversion where coarse-graining gives $D_l\St/Dt = 0$ identically since every inviscid term in the budget includes a subfilter moment with $u_i$ in one slot; this complicates interpretation of the inner-domain boundary flux terms as cross-scale exchanges. \citet{craske_transport_2025} show the underlying framework admits material as well as fixed control volumes, which would remove this particular issue at the expense of subdomains that deform with the flow. The more basic limitation is that the inner APE, defined by sorting within a subdomain, vanishes as the subdomain's vertical extent goes to 0 --- a case relevant to many geophysical settings, and the limit in which the coarse-graining is simplest (\S\ref{sec:horizontal}). A hybrid approach, sorting in the vertical while filtering in the horizontal, may be worth exploring. 

\section{Summary and Conclusions}\label{sec:conclusions}
In this work we develop a coarse-graining (filtering) approach for the local APE density. A scale decomposition appropriate for filters with both vertical and horizontal extent is accomplished by referencing the APE of the filtered buoyancy field to the filtered reference profile (\ref{eq:split}), such that the small-scale APE is the energy remaining to be released by rearrangement of the subfilter parcels after release of the large-scale APE. This decomposition separates the total APE into two positive semi-definite energies, both of which vanish for a fluid at rest, and which can be composed from a series of band-passed energies so that contributions at different scales can be meaningfully compared. Evolution equations for both large- and small-scale APE are derived, including the exact form of the cross-scale APE transfer term, a counterpart to the coarse-grained KE equations \citep{eyink_localness_2009, aluie_mapping_2018}. In the limit of horizontal filtering the scale decomposition simplifies, extending the Reynolds-averaged decompositions of \citet{scotti_diagnosing_2014} and \citet{tailleux_energetically_2025}. This coarse-graining approach can be contrasted to prior scale-resolved treatments of potential energy in stratified flows which generally have relied on approximate quadratic forms of APE which assume small-parcel displacements around a fixed-reference profile \citep{lindborg_energy_2006, augier_spectral_2012, zhang_analysis_2022, foldes_characterization_2025} --- assumptions violated in many classes of flows --- or used spectral decomposition thereby losing spatial-locality \citep{molemaker_local_2010}.

The coarse-graining approach is illustrated with a simulation of three-dimensional Kelvin--Helmholtz instability at $Re = 1032$, $Ri = 0.1$, $Pr=1$. Prior work on the spatially-local energetics of this flow has, to our knowledge, not decomposed the APE by scale, and where decompositions of the KE have been used they have been separations into mean, two-dimensional and three-dimensional reservoirs \citep{caulfield_anatomy_2000, smyth_turbulence_2003, liu_effects_2023, yi_energy_2026}, that is, projections onto the spanwise mean rather than separations by length scale. The coarse-graining approach developed here instead resolves the exact energetics continuously in scale, with two consequences for the picture of the instability. First, the scale-resolved budgets track the route that energy takes during the initial two-dimensional phase: KE is transferred forward at the billow scale, converted to APE, and then transferred further forward as APE. A portion of this small-scale APE is then converted back to KE and transferred upscale, describing a loop in scale space noted in other stratified two-dimensional flows \citep{boffetta_flux_2011, zhou_scale--scale_2016, zhao_scale_2022}, and inferred previously for KH from the vorticity dynamics rather than shown directly from the scale-dependent energetics \citep{staquet_two-dimensional_1995}.

Second, the budgets separate the transfer of APE to small scales (stirring) from its dissipation (mixing), in both time and space. Forward APE transfer has largely subsided before small-scale APE dissipation peaks (here near $t\approx70$). Following this time the conversion of APE to KE at $\ell=1$ is small in the volume integral \citep{caulfield_anatomy_2000}, although this conceals large dipoles of opposite-signed conversion around the billow core which persist past the onset of the secondary spanwise instabilities ($t\approx 90$) at similar spatial scale. During this period APE dissipation drops, but then remains relatively steady compared to the time-variability in its source terms of cross-scale transfers and KE-APE conversions. Forward KE transfers are enhanced at the stagnation points, where braid instabilities are known to originate \citep{caulfield_anatomy_2000, mashayek_zoo_2012}, and grow in time in the transition to three-dimensional turbulence. This is tracked by an increase in KE dissipation, which at this scale remains localized in space along the billow eyelids with little spanwise structure relative to the KE production terms.
 In contrast, the APE dissipation is consistently intensified in the braids, particularly where they meet the billow, indicating regions of persistent locally efficient mixing even when the bulk energetics favor dissipation of KE. There is little diabatic mixing in the billow core itself at any stage (at a scale of $\ell=1$), despite regions of static instability, consistent with observations and the local scalar-variance dissipation rate in simulations at higher Reynolds number \citep{geyer_mixing_2010, lewin_influence_2021, lefauve_structure_2026}.

The power of this approach lies in the ability to simultaneously decompose the flow by spatial scale while retaining information about the spatial locality of terms such as the cross-scale energy fluxes. Previous work has used the coarse-graining approach with KE to investigate the mechanisms responsible for cross-scale KE transfers in stratified flows ranging in scale from three-dimensional turbulence to large-scale ocean circulation, and the work presented here enables a more complete consideration of a full energy cycle. Potential applications suggested by the KH example above include the use of anisotropic filters as a scale-resolved counterpart to energy decompositions by dimensionality \citep{smyth_turbulence_2003, liu_effects_2023, yi_energy_2026}, investigation of scale- and space-resolved mixing efficiencies \citep{lewin_influence_2021, couchman_mixing_2023, tailleux_energetically_2025}, or the development of mixing parameterizations based on the cross-scale APE flux term $\Pi_A$ (\ref{eq:Pi_A}). Given the fundamental role that APE plays in the energetics of stratified flows --- including the direct connection between the dissipation of APE and diabatic mixing --- we anticipate that the framework introduced here may provide new insight into processes at many spatial scales. 

\backsection[Supplementary data]{\label{SupMat}Supplementary material and animations are available at \textit{TBD on publication}.}

\backsection[Acknowledgements]{Helpful comments from John Craske, Remi Tailleux, and an anonymous reviewer are gratefully acknowledged. We also thank Alexis Kaminski and Zhihua Zheng for discussions during manuscript preparation. Claude Code, Opus 5, and Fable 5.1 were used for portions of the software development, assisting with derivations, drafting the schematic figures, and for suggesting edits during revision of the text. All output has been reviewed and verified by the authors.}

\backsection[Funding]{We gratefully acknowledge support from the NSF OCE-2446313 (J.O.W. and T.C.), OCE-2342990 (J.O.W.), and ONR N000142412583 (J.O.W.). R. B. acknowledges support from the Israel Science Foundation grant 2054/23, the ERC (ML
Transport, 101163887), and NSF grant OCE-2446687.}

\backsection[Declaration of interests]{The authors report no conflict of interest.}

\backsection[Data availability statement]{The code to generate the numerical simulation data and analyses used in this study is available at https://github.com/tomchor/CoarseGrainedKHAPE/releases/tag/v1.0.}

\appendix
\section{Scale hierarchy}
\label{sec:hierarchy}

The scale decomposition (\ref{eq:split}) can be applied to any pair of buoyancy field $\beta(\boldsymbol{x},t)$ and monotonically increasing reference profile $\beta_*$,
\begin{equation}
\overline{E_A(\beta,z,t;\beta_*)}
 = E_A\big(\overline{\beta},z,t;\langle\beta_*\rangle\big)
 + \overline{E_A\big(\beta,z+\widetilde{\Upsilon}_{\beta}(\boldsymbol{x}, t),\,t;\beta_*\big)},
\label{eq:general}
\end{equation}
where $\widetilde{\Upsilon}_{\beta}=\widetilde{\zeta}_\beta - z$ with $\widetilde{\zeta}_\beta$ the resting level of $\overline{\beta}$ in the $\langle\beta_*\rangle$ profile. In (\ref{eq:split}) this decomposition is applied to the unfiltered fields $(b,b_*)$; however, the same decomposition can also be applied to the coarse fields themselves, $(\overline{b},\langle b_*\rangle)$, allowing the decomposition to be composed from a series of band-passed contributions.

Let $G_1,G_2,\dots$ be filter kernels (subject to the same constraints as in \S\ref{sec:SpatialFiltering}) with scales $\ell_1<\ell_2<\dots$. These are nested in the
sense that they can be written as $G_n=H_n*G_{n-1}$, where $H_n$ is again a non-negative kernel and $G_0=\delta$. A field filtered by $G_n$ is written
$\overline{f}^{\,n}$, with corresponding filtered reference profile $\langle b_*\rangle_n$, and a single application of the increment $H_n$ is
written $\overline{(\,\cdot\,)}^{\,H_n}$. The filtered state at scale $\ell_n$ is the pair $(\overline{b}^{\,n},\langle b_*\rangle_n)$, which has large and small-scale APE
\begin{equation}
L_n \equiv E_A\big(\overline{b}^{\,n},z,t;\langle b_*\rangle_n\big),
\qquad
S_n \equiv \overline{E_A}^{\,n}-L_n,
\label{eq:LnSn}
\end{equation}
so that for application of a single filter $L_1=E_A^l$ and $S_1=E_A^s$ of (\ref{eq:split}).

Now take the coarse state at scale $\ell_{n-1}$ and filter it with the increment $H_n$. Since $L_{n-1}$ is itself an APE, formed from the $(n-1)$-filtered state, (\ref{eq:general}) gives
\begin{equation}
\overline{L_{n-1}}^{\,H_n}
 = \underbrace{E_A\big(\overline{b}^{\,n},z,t;\langle b_*\rangle_n\big)}_{L_n}
 + \underbrace{\overline{E_A\big(\overline{b}^{\,n-1},z+\widetilde{\Upsilon}_n(\boldsymbol{x},t),\,t;
   \langle b_*\rangle_{n-1}\big)}^{\,H_n}}_{B_n},
\label{eq:step}
\end{equation}
where $\widetilde{\Upsilon}_n = \tilde\zeta_n - z$ with $\tilde\zeta_n$ the
resting level of the scale-$\ell_n$ filtered parcel, and we have used the fact
that applying $H_n$ to $\overline{b}^{\,n-1}$ gives $\overline{b}^{\,n}$,
and likewise $\langle b_*\rangle_{n-1}$ filters to $\langle b_*\rangle_n$.
The first term on the right-hand side is identically $L_n$, the large-scale
APE at scale $\ell_n$. However, filtering $L_{n-1}$ differs from this by the APE held in the band of scales between $\ell_{n-1}$ and $\ell_n$ (denoted $B_n$). This term can be understood as the second stage of the two-stage process of \S\ref{sec:physical} applied to the $\ell_{n-1}$-scale filtered state rather than the fluid itself. It is an average of a positive semi-definite quantity, hence $B_n\ge0$.

Filtering $\overline{E_A}^{\,n-1}=L_{n-1}+S_{n-1}$ with $H_n$ gives $\overline{E_A}^{\,n}=\overline{L_{n-1}}^{\,H_n}+\overline{S_{n-1}}^{\,H_n}$,
and using (\ref{eq:step}) it follows that
\begin{equation}
S_n = \overline{S_{n-1}}^{\,H_n} + B_n .
\label{eq:Srec}
\end{equation} 
Each filtering step thus moves the band energy of the increment from the large-scale to the small-scale reservoir, and iterating (\ref{eq:Srec}) from $S_0=0$ gives, at every point,
\begin{equation}\label{eq:EABands}
    \overline{E_A}^n = L_n + \sum_{m=1}^n \overline{B_m}^{H_n*\cdots*H_{m+1}},
\end{equation}
the APE at scale $\ell_n$ decomposed into the large-scale APE plus the band energies at all finer scales, each smoothed by any increments applied after it. Every term is non-negative and there are no cross-band terms, the analog for APE of the successive filtering-band kinetic energies of \citet{eyink_localness_2009}. On integrating over the domain the filtering in (\ref{eq:EABands}) drops out, so that $\int E_A\,\mathrm{d}V = \int L_n\,\mathrm{d}V + \sum_{m=1}^n \int B_m\,\mathrm{d}V$ up to contributions of the boundary extension from within approximately a filter width of the walls (\S\ref{sec:split}), which vanish when the fluid there is at rest.

We note that the alternative of using the unfiltered reference profile, as discussed in \S\ref{sec:physical}, fails to admit a scale hierarchy that composes in this manner. Defining the subgrid energy as a remainder gives sign-indefinite band terms, whereas the three-term split has non-negative energies in each band, but each band also requires the sign indefinite term $C_n$ which is not an energy.

\bibliographystyle{jfm}
\bibliography{jacob}

\end{document}